\documentclass[12pt]{article}
\usepackage[T2A]{fontenc}
\usepackage[cp1251]{inputenc}
\usepackage[english]{babel}
\usepackage[dvips]{graphicx}
\normalsize
\usepackage{amssymb,amsmath}

\sloppypar

\def\vt{$\xi_{\rm t}$}

\newcommand{\Teff}{T_{\rm eff}}
\newcommand{\eps}{\log\varepsilon}
\newcommand{\kms}{km\,s$^{-1}$}

\newcommand{\kH }{$S_{\rm H}$}    

\begin{document}
\selectlanguage{english} 
\title{Influence of Inelastic Collisions with Hydrogen Atoms on Non-LTE	Oxygen Abundance Determination \\ in the Sun and late-type stars}

\author{T. ~M. ~Sitnova \thanks{sitnova@inasan.ru}\;\;$^{1,2}$\; L. ~I.~Mashonkina$^1$\\[2mm]
\begin{tabular}{l}
 $^1$ {\it Institute of Astronomy, Russian Academy of Sciences, Moscow, Russia}\\[2mm]
 $^2$ {\it Herzen State Pedagogical University, St. Petersburg, Russia}
\end{tabular}
}
\date{}
\maketitle

\begin{abstract}
We present the non-local thermodynamic equilibrium (non-LTE) calculations for O~I with the updated model atom that includes quantum-mechanical rate coefficients for O~I~+~H~I inelastic collisions from the recent study of Barklem (2018). The non-LTE abundances from the O~I lines were determined for the Sun and 46 FG stars in a wide metallicity range, $-2.6 <$ [Fe/H] $< 0.2$. An application of accurate atomic data leads to larger departures from LTE and lower oxygen abundances compared to that for the Drawin's theoretical approximation. For the infrared O~I 7771-5~\AA\ triplet lines, the change in the non-LTE abundance is $-0.11$~dex for the solar atmospheric parameters and decreases in absolute value towards lower metallicity. We revised the [O/Fe]--[Fe/H] trend derived in our earlier study. The change in [O/Fe] is small in the [Fe/H] range from --1.5 to 0.2. For stars with [Fe/H] < -1, the [O/Fe] ratio has increased such that [O/Fe] = 0.60 at [Fe/H] = --0.8 and increases up to [O/Fe] = 0.75 at [Fe/H] = --2.6.

\end{abstract}

\maketitle

\section{Introduction}
The oxygen abundance in the atmospheres of the Sun and stars is an important quantity for testing the Galactic chemical evolution scenarios and the theory of stellar structure and evolution.

The infrared (IR) O~I 7771-5 \AA\  triplet lines
are the only set of atomic lines  observed
in  spectra of metal-poor stars. 
Previously, many authors have shown that the IR O~I lines are formed under conditions far from local thermodynamic equilibrium (LTE). 
The oxygen NLTE abundance
was first determined by Kodaira
and Tanaka (1972) and Johnson (1974) for stars and
by Shchukina (1987) for the Sun. Subsequently,
more comprehensive oxygen model atoms were constructed
by Kiselman (1991), Carlsson and Judge (1993),
Takeda (1992), Paunzen et al. (1999), Reetz (1999),
Mishenina et al. (2000), and Przybilla et al. (2000).
Non-LTE leads to a strengthening of O~I IR lines and,
consequently, to a decrease in the abundance derived
from these lines.

Having considered atomic and molecular lines in
the solar spectrum, Asplund et al. (2004) achieved
agreement between the abundances from different
lines using a three-dimensional (3D) model atmosphere based on hydrodynamic calculations and the
non-LTE corrections calculated with a classical 1D
model atmosphere. In Asplund et al. (2004), the mean
abundance from atomic and molecular lines is  $\eps$ =
8.66 $\pm$ 0.05\footnote{Here,  $\eps$ = log(N$_{\rm elem}$/N$_{\rm H}$) + 12.}; further, Asplund et al. (2009)
obtained  $\eps$ = 8.69 by the same method.  This value turned out to
be lower than $\eps$ = 8.93 $\pm$ 0.04 obtained previously
by Anders and Grevesse (1989) from OH molecular
lines using the semi-empirical HM74 model atmosphere (Holweger and Mueller 1974). It is worth noted that the models of solar internal structure constructed with the chemical composition from Anders
and Grevesse (1989) described well the sound speed
and density profiles inferred from helioseismological
observations. A revision of the oxygen abundance by
0.27~dex led to a discrepancy between the theory and
observations up to 15$\sigma$ (Bahcall and Serenelli 2005).
This problem still remains unsolved.

In our previous paper (Sitnova et al. 2013), based
on the model atom from Przybilla et al. (2000) improved by including 
rate coefficients for electron-impact excitation of O~I
from Barklem (2007), we derived the mean oxygen
abundance $\eps$ = 8.74 $\pm$ 0.05 from the O~I 6300,
6158, 7771-5, and 8446 \AA\ lines in the solar spectrum
using a classical plane-parallel solar model atmosphere and $\eps_{\rm +3D}$ = 8.78 $\pm$ 0.03 by applying the
3D corrections from the Caffau et al. (2008).

There is a huge number of studies where the
[O/Fe] ratio was determined for samples of stars;
here, we mention some of them. It is well known from
observations that stars with [Fe/H]\footnote{We use the standard notation for the elemental abundance
	ratios [X/H] = log(N$_{\rm X}$/N$_{\rm H}$)$_{\rm star}$ -- log(N$_{\rm X}$/N$_{\rm H}$)$_{\rm Sun}$} <
0 show positive [O/Fe] ratio, which increases with
decreasing metallicity approximately down to [Fe/H] $\simeq$ --1
and then remains almost constant at [Fe/H] < --1.
The [O/Fe] ratio as a function of [Fe/H] is well-understood qualitatively and quantitatively, and, in
general, the Galactic chemical evolution models describe the observational data for [O/Fe]. Almost the
same [O/Fe] ratio for stars with [Fe/H] < --1 stems
from the fact that, at the epoch of their formation, the
interstellar gas was enriched with metals by massive
stars exploded as type II supernovae (SNe~II) or
hypernovae. By the formation epoch of stars with
[Fe/H] $\simeq$ --1, type Ia supernovae began to contribute
to the enrichment of the medium with heavy elements,
where the iron production efficiency is higher than
that in the explosions of massive stars, which led to a
decrease in [O/Fe]. At present, attempts are being
made to establish subtle features in the behavior of
[O/Fe], for example, to understand what the actual
spread in [O/Fe] is for stars with a similar metallicity,
from which the conclusion about the mixing of matter
in the Galaxy can be drawn. Such an attempt was
made by Bertran de Lis et al. (2016), who determined
the oxygen abundance from IR OH lines in red giants
with metallicity --0.65 < [Fe/H] < 0.25. Ram{\'{\i}}rez
et al. (2013) determined the oxygen abundance by
taking into account the departures from LTE from the
O~I 7771--5 \AA\ lines for a sample of several hundred
FGK dwarfs with --1.2 < [Fe/H] < 0.4. Bensby
et al. (2014) performed a detailed analysis of 13 elements from oxygen to barium in several hundred
nearby dwarf stars with --2.6 < [Fe/H] < 0.4. The
oxygen abundance was inferred from the O~I 7771--5 \AA\ lines in non-LTE. In these papers, particular
attention is given to the chemical peculiarities of
stars in various Galactic subsystems (the thin and
thick disks, the halo, the Hercules and Arcturus
streams). Amarsi et al. (2015) collected data from the
literature on oxygen abundance determinations over
2000--2015 for dwarf stars with --3.3 < [Fe/H] <
0.5, redetermined their atmospheric parameters using
the temperatures from the IR flux method,
and corrected the derived abundance by applying
the corrections for the hydrodynamic and non-LTE
effects. As a result, the authors found a linear increase
in [O/Fe] from --0.3 to 0.6 as [Fe/H] decreased
from 0.5 to --0.7 and then a constant [O/Fe] down
to [Fe/H] $\simeq$ --2.5; with a possible increase in [O/Fe] to 0.8 
 at lower metallicity. The question of
whether there is an increase in [O/Fe] at [Fe/H] < --1
is not completely clear. There are different opinions
on that score in the literature. For example, from analysis of
OH lines, Israelian et al. (1998) found a linear increase in [O/Fe] from 0.6
to 1 for [Fe/H] from --1.5 to --3 . Fulbright and Johnson (2003) also found
an increase in [O/Fe] from 0.6 to 0.8 as [Fe/H] decreased from --1 to --2.5 using the forbidden 6300~\AA\ 
line and the IR 7771-5 \AA\ triplet lines. However, the
results of Tomkin et al. (1992), Gratton et al. (2000),
Nissen et al. (2002), Cayrel et al. (2004), Fabbian
et al. (2009), Sitnova (2016),  and Zhao et al. (2016), which are also
based on an analysis of atomic O~I lines, suggest a
constant [O/Fe] ratio at [Fe/H] < --1.

For the Sun and cool stars, the non-LTE abundance derived from the atomic lines can be inaccurate due to the
uncertainty in  the statistical equilibrium calculations
 related to the lack of knowledge of the excitation and ionization efficiencies in inelastic collisions with neutral hydrogen. 
Quantum-mechanical
calculations of collisions with hydrogen atoms are
available for a number of atoms: for Li~I (Belyaev
and Barklem 2003), Na~I (Belyaev et al. 1999, 2010;
Barklem et al. 2010), Mg~I (Barklem et al. 2012),
Al~I (Belyaev 2013), Si~I (Belyaev et al. 2014), K~I
(Belyaev and Yakovleva 2017), and Ca~I (Belyaev
et al. 2016; Barklem 2016; Mitrushchenkov et al.
2017). For those atoms, where  accurate calculations or
laboratory measurements are missing, the rates of inelastic collisions with H~I are calculated from the formula derived by Steenbock and Holweger (1984) using the
formalism of Drawin (1968, 1969). The authors
themselves estimate the accuracy of the formula to
be one order of magnitude. A scaling factor (\kH)
that can be found empirically by reconciling the abundances from
lines with strong and weak departures from LTE is
usually introduced in this formula. For oxygen, Allende Prieto et al. (2004) and Pereira et al. (2009)
obtained \kH\ = 1 by investigating the changes of the
center-to-limb variation of O~I line profiles in solar intensity spectrum;
Takeda (1995) obtained the same result from O~I lines
in the solar spectrum in fluxes. For the Sun and cool
stars, agreement between the abundances determined
from different O~I lines can be achieved by choosing
the scaling factor. Caffau et al. (2008) performed
non-LTE calculations for O~I with \kH\ = 0, 1/3, and
1. The higher the efficiency of collisions with
hydrogen atoms leads to the smaller departures from LTE
and the higher abundance. For example, for the IR
O~I 7771 \AA\ line the abundance derived with \kH\ = 1
is higher than that with \kH\ = 0 by 0.12 dex. It is
important to note that Belyaev and Yakovleva (2017)
proposed a simplified but physically realistic method
of estimating the rates of inelastic collisions with
hydrogen atoms that is recommended to be applied
instead of Drawin's approximation for those elements
where accurate data are not available so far.

This study was motivated by the appearance of
quantum-mechanical rate coefficients for O~I + H~I inelastic collisions performed by Barklem (2018).
In this paper, we check how the use of data from
Barklem (2018) affects the oxygen abundance
determination for the Sun and a sample of FG stars
with metallicity --2.6 < [Fe/H] < 0.2 and on the
[O/Fe]--[Fe/H] trend derived in our earlier study
(Zhao et al. 2016). The model atom and the methods
and codes used are described in Section 2. The O~I
lines in the Sun are analyzed in Section 3. Stellar atmospheric
parameters, observations, and the derived oxygen
abundance are described in Section 4. Our results
are presented in the Conclusions.

\section{Method of oxygen abundance determination}\label{method}

We determined the oxygen abundance from O~I lines in the non-LTE case where the
population of each level in a model atom
is calculated by simultaneously solving the system
of statistical equilibrium (SE) and radiative transfer
equations. To solve this system of equations in a
specified model atmosphere, we use the DETAIL code
developed by Butler and Giddings (1985) based on
the accelerated $\Lambda$-iteration method. The opacity calculation was improved, as described by Mashonkina
et al. (2011). The level populations obtained in DETAIL were then used to compute the line profiles with
the synthV\_NLTE code (Tsymbal 1996, updated in Ryabchikova et al. 2016).
We use O. Kochukhov's binmag\footnote{http://www.astro.uu.se/$\simeq$oleg/download.html}
code to fit the theoretical spectrum with the
observed one.

The technique of our calculations and the mechanism
of departures from LTE for O~I were described in Sitnova et al. (2013). We use the multilevel model atom constructed from the most up-to-date atomic data. We adopted the model atom from
Przybilla et al. (2000) as a basis; it consists of 51
O~I levels and the O~II ground state. The level
energies were taken from NIST (Kramida et al. 2015);
the transition oscillator strengths and photoionization cross sections were taken from the Opacity
Project (Seaton et al. 1994), which are accessible in
the TOPbase\footnote{http://cdsweb.u-strasbg.fr/topbase/topbase.html} database. To calculate the rates of
bound-bound transitions in collisions with electrons,
we use the quantum-mechanical calculations from
Barklem (2007) for 153 transitions. For the remaining transitions, where there are no accurate data,
we use the formulas from van Regemorter (1962)
and Wooley and Allen (1948) for optically allowed
and forbidden transitions, respectively. To calculate
the rates of bound-free transitions in collisions with
electrons, we use the formula from Seaton (1962)
with the threshold photoionization cross section from
TOPbase. The resonant charge exchange (O~I + p $\leftrightarrow$
O~II + H I) was taken into account as prescribed by
Arnaud and Rothenflug (1985).
In this study, we take into account the excitation/deexcitation and ion-pair formation/mutual
neutralization processes in collisions with hydrogen
atoms according to the data from Barklem (2018).
Previously, we used Drawin's approximation (Drawin
1968, 1969; Steenbock and Holweger 1984) due to
the absence of accurate data.
Figure~1 shows the transition rates in inelastic
collisions with hydrogen atoms and electrons under
conditions typical for the solar atmosphere at the
O~I line formation depths. According to the data
from Barklem (2018), the O~I + H~I excitation rates
are systematically lower than those calculated from
Drawin's formula. On average, the difference between the rates is about two orders of magnitude. For
transitions with energies $\Delta$E = E$_u$ -- E$_l$ < 1.5 eV
Barklem's rates for O~I + H I collisions are of the
same order of magnitude as the rates of collisions with
electrons, while, for transitions with higher energies,
the electron collisions are much more efficient than
the hydrogen ones. 
An ion pair production in collisions with H~I  is  much more efficient than an
ionization in collisions with electrons in the entire range of energies. 
However, this does not affect the results, because the statistical equilibrium
of O~I in the range of parameters under consideration is determined by the bound-bound transitions.
We performed a test calculation for the Sun 
neglecting ion-pair formation and mutual neutralization processes in collisions with hydrogen atoms
and obtained very small changes in level populations that led to changes in the non-LTE abundance
within 0.003~dex for the O~I 7771-5 \AA\ lines.

\begin{figure}[htbp]
	\resizebox{85mm}{!}{{\includegraphics{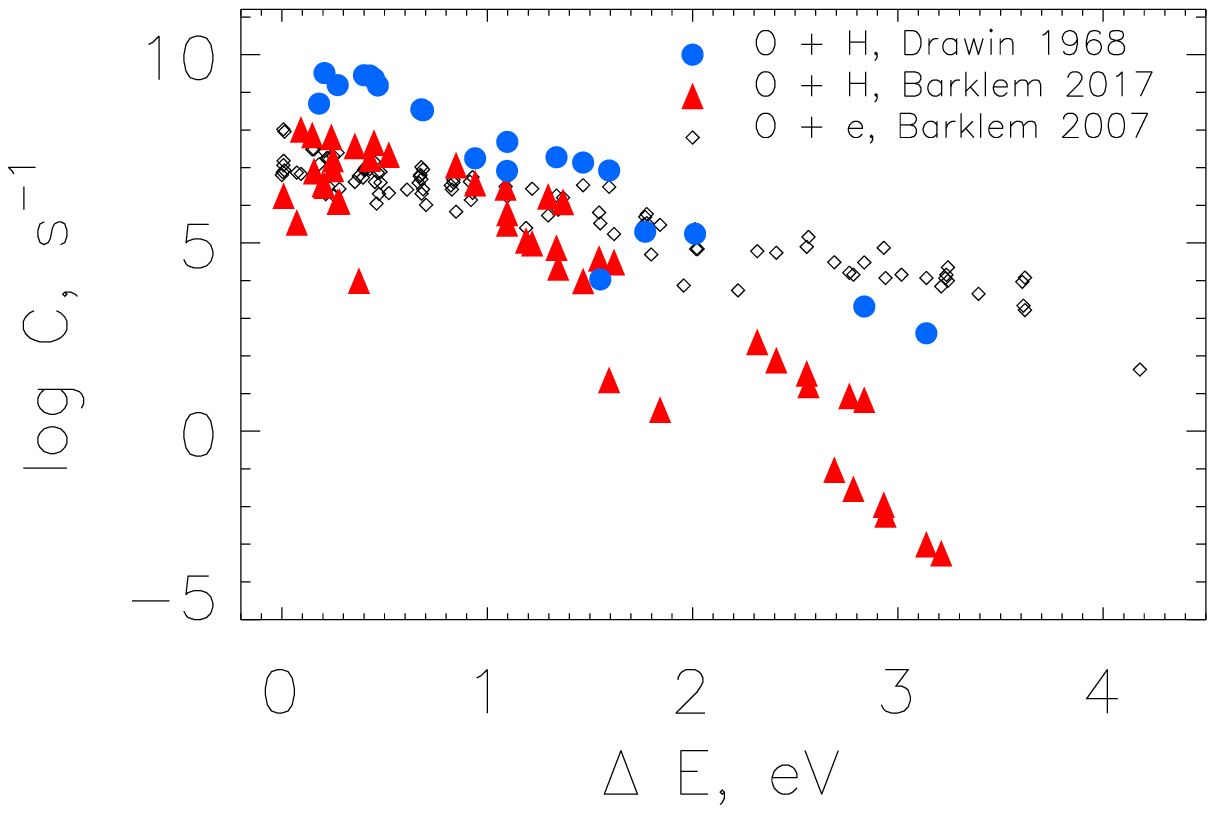}}}
	\resizebox{85mm}{!}{{\includegraphics{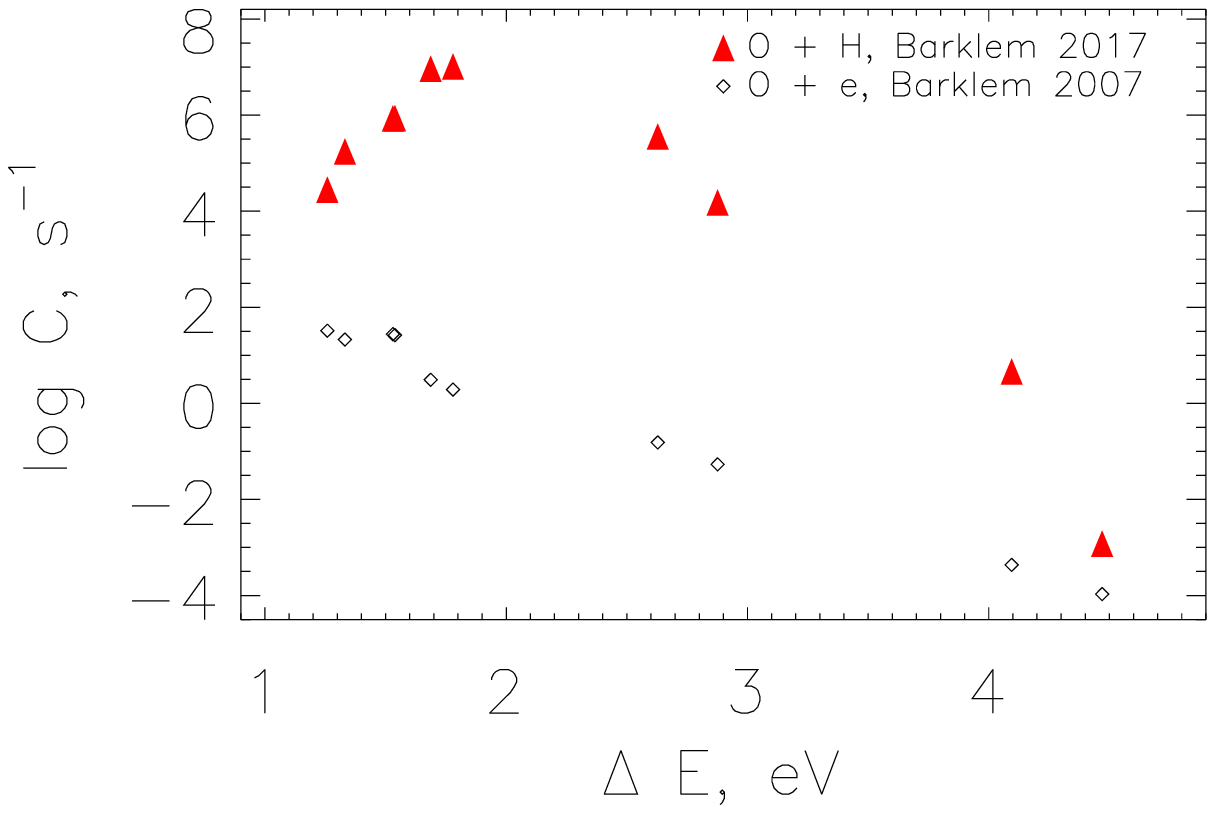}}}
	\caption{
Left panel: Excitation rates in collisions with electrons (diamonds) and hydrogen atoms, according to the
quantum-mechanical calculations of Barklem (2018) (triangles) and the approximation of Drawin (1968) (circles), versus
transition energy. Right panel: The rates of the processes O I + e$^-$ $\leftrightarrow$ O II + 2e$^-$ (diamonds) and O I + H I $\leftrightarrow$ O II + H$^-$ (triangles).
The rates were calculated for the density of hydrogen atoms log N$_H$ = 17, the electron density log N$_e$ = 14, and the
temperature T = 6840 K.		
		} 
	\label{rates}
\end{figure}

\section{Analysis of O~I lines in the solar spectrum} 
\label{sun}

The solar abundance was determined using the
spectrum of the Sun as a star (Kurucz et al. 1984).
The model atmosphere has an effective temperature
 $\Teff$ = 5780~K, surface gravity log g = 4.44, and microturbulent velocity \vt = 0.9 \kms. We use the classical 1D models from the MARCS grid (Gustafsson
et al. 2008).

The list of lines and the derived oxygen abundance
are given in Table~1. The atomic data for transitions,
i.e., the wavelength $\lambda$, the oscillator strength (log~gf),
and excitation energy of  the lower level (E$_{\rm exc}$), were
taken from the VALD database (Kupka et al. 1999;
Ryabchikova et al. 2015).

The application of accurate data for collisions with
hydrogen atoms led to an increase in the departures from LTE and a strengthening of the IR lines.
We obtained the mean non-LTE oxygen abundance
 $\eps$(O) = 8.69 $\pm$ 0.08, which is lower than that derived in non-LTE with Drawinian rates by 0.07 dex.
For the IR triplet lines in the Sun, the non-LTE abundance decreased by 0.11 dex. For the 8446 \AA\ lines the
change is slightly smaller, 0.07 dex. In comparison
with our previous results (Sitnova et al. 2013), the
difference between the non-LTE abundances from
the 7771-5 and 8446 \AA\ lines increased from 0.02 to
0.05 dex. For the weak 6158 and 6300 \AA\ lines in
the visible range, the departures from LTE are still
negligible, irrespective of the method of calculating
the inelastic collisions with hydrogen atoms. The
difference of 0.07~dex between the non-LTE abundances from the forbidden [O~I] 6300 \AA\ line and
the 7771-5 \AA\ lines obtained in our previous paper
decreased to 0.04 dex. The difference in non-LTE
abundances from the 6158 and 7771-5 \AA\ lines increased from 0.07 to 0.18 dex. It should be noted
that the abundance from the 6158 \AA\ line is systematically higher than that from the IR triplet not only
for the Sun but also for FG dwarfs. According to the
data from Zhao et al. (2016), the mean difference $\Delta$ =
0.12 $\pm$ 0.07 is obtained by comparing the absolute
non-LTE abundances from these two lines for 17
FG stars in the [Fe/H] range from --0.8 to 0.2. With
the classical MARCS model atmosphere Asplund
et al. (2004) obtained a difference of 0.13~dex in the
non-LTE abundance between these two lines for the
Sun. The systematic difference is probably caused by
the inaccuracy of the oscillator strengths calculated
theoretically by Hibbert et al. (1991).

It is worth noting that using 3D non-LTE line formation calculations with O~I + H~I rate coefficients  from Barklem (2018), Amarsi et al. (2018) derived  $\eps$(O) = 8.69 $\pm$ 0.03 from the O~I 7771--5 \AA\ lines. Our NLTE modelling with classical 1D model atmosphere gives average non-LTE abundance from the IR triplet lines, $\eps$(O) = 8.65 $\pm$ 0.03, which is consistent within the error bars with those derived by Amarsi et al. (2018).

\begin{table}[htbp]
	\caption{
The list of O I lines with atomic parameters and LTE and non-LTE solar abundances. The non-LTE abundance
is given for the cases where the inelastic collisions with hydrogen atoms were calculated from the approximate formula
(S13) and based on the quantum-mechanical calculations (S17)		
		}
	\label{lines}
	\tabcolsep3.7mm
	\begin{center}
		\begin{tabular}{llcccccrrr}
			\hline
			$\lambda$, \AA   & E$_{exc}$, eV   & log gf &   transition & LTE & non-LTE, S13 & non-LTE, S17 \\    
			\hline
			6158.146   & 10.741 & --1.841  &  $3p\; ^5P$  -- $4d\; ^5D^\circ$ & 8.83 & 8.82 & 8.82 \\
			6158.176   & 10.741 & --0.996  &  $3p\; ^5P$  -- $4d\; ^5D^\circ$ &  8.83 & 8.82 & 8.82 \\
			6158.186   & 10.741 & --0.409  &  $3p\; ^5P$  -- $4d\; ^5D^\circ$ &  8.83 & 8.82 & 8.82 \\
			6300.304   &  0.000 & --9.720$^1$  &  $2p\; ^3P$ -- $2p\; ^1D$ & 8.68 & 8.68 & 8.68 \\
			7771.941   &  9.146 &  0.369  &  $3s\; ^5S^\circ$  -- $3p\; ^5P$ & 8.91 & 8.74 & 8.63 \\
			7774.161   &  9.146 &  0.223  &  $3s\; ^5S^\circ$  -- $3p\; ^5P$ & 8.91 & 8.76 & 8.65 \\
			7775.390   &  9.146 &  0.001  & $3s\; ^5S^\circ$  -- $3p\; ^5P$ & 8.87 & 8.75 & 8.66 \\
			8446.250   &  9.521 &  --0.463 & $3s\; ^3S^\circ$  -- $3p\; ^3P$ & 8.87 & 8.77 & 8.70 \\			
			
			\hline
		\end{tabular}
	\end{center}
	$1$ -- The data from Froese Fisher et al. (1998). \\
	When fitting the [O I] 6300 \AA\ line, we took into account the blending Ni~I 6300.34 line with log(gf) = --2.11 (Johansson et al. 2003)
	and the abundance $\eps$ (Ni) = 6.23.
\end{table}  %

\section{Oxygen abundance in the sample stars} 
\label{stars}

In this section, we redetermine the non-LTE oxygen abundance based on an improved model atom
for 46 FG stars investigated by us previously (Zhao et al. 2016).

\subsection{Stellar sample, observations, and atmospheric parameters} 
\label{obspar}

The sample of stars includes 46 unevolved stars,
from dwarfs to subgiants. The stars are uniformly
distributed in metallicity over a wide range, --2.6 <
[Fe/H] < 0.2.
High-resolution ($\lambda$/$\Delta$$\lambda$ > 45~000)
spectra with a signal-to-noise ratio S/N > 60 were
obtained at the 3-m telescope of the Lick Observatory
with the Hamilton spectrograph or taken from
the UVES\footnote{http://archive.eso.org/eso/eso\_archive\_main.html} and ESPaDOnS\footnote{http://www.cadc-ccda.hia-iha.nrc-cnrc.gc.ca/en/search/} archives. We also
used the spectra obtained at the 2.2-m telescope
of the Calar Alto Observatory with the FOCES
spectrograph and provided by K. Fuhrmann. The observations and their reduction were described in detail
by Sitnova et al. (2015), Pakhomov and Zhao (2013),
and Zhao et al. (2016).

We use the atmospheric parameters carefully
investigated by various methods in Sitnova et al.
(2015). For the majority of stars, there are effective temperature determinations by the infrared flux method
(Alonso 1996; Casagrande et al. 2010, 2011); the
effective temperature for each star can be estimated
from its color indices. The adopted temperatures agree
with the photometric ones within the errors of determination
 and were chosen so as to provide agreement between the non-LTE abundances from Fe~I
and Fe II lines.  For ten
stars of our sample, there are $\Teff$ determinations
based on the bolometric fluxex and the angular diameters
measured with the CHARA interferometer (Boyajian
et al. 2012, 2013; North et al. 2009; von Braun
et al. 2014). Our temperatures are systematically
higher than the interferometric ones by 78 $\pm$ 81~K.
However, this difference does not exceed the error in
$\Teff$, which we estimate to be 80 K. For two other
stars, HD~140283 and HD~103095, the interferometric temperatures measured previously by Creevey
et al. (2012, 2015) and Boyajian et al. (2013) differ
from those adopted by us by more than 250 K.
Karovicova et al. (2018) redetermined the angular
diameters and bolometric fluxes for these two stars,
which led to an increase in $\Teff$ and agreement with
our determinations within 10~K.

The Hipparcos trigonometric parallaxes
(van Leeuwen et al. 2007) are available for all sample stars, which allows to calculate log~g by a
spectroscopy-independent method. For the stars
where the Hipparcos parallax error exceeds 10\%,  surface gravities were refined by analyzing
the non-LTE abundances from Fe~I and Fe~II lines.
The statistical equilibrium of Fe~I--II was calculated
with the model atom developed by Mashonkina
et al. (2011) with the scaling factor to the Drawin
rates of inelastic collisions with hydrogen atoms \kH\ =
0.5. It is worth noting that, in the first Gaia\footnote{https://www.cosmos.esa.int/gaia} data release (DR1),  parallaxes for 22 stars of
our sample became available (Brown et al. 2016). We achieved good agreement between
our spectroscopic log~g and those calculated from the
 Gaia DR1 parallaxes: $\Delta$log~g(Spec -- Gaia) =
--0.01 $\pm$ 0.07, despite the fact that for the same
22 stars the spectroscopic log~g differ noticeably from
those derived from the Hipparcos data, $\Delta$log~g(Spec --
Hipparcos) = --0.15 $\pm$ 0.12. The differences between log~g$_{\rm spec}$ and log~g$_{\rm Hipp}$
concern the stars that are 100~pc or more away from
the Sun. For the nearest stars, the spectroscopic log~g
agree well with those calculated from both Hipparcos
and Gaia~DR1 data.

The microturbulent velocity \vt\ and [Fe/H] were derived
from Fe~I and Fe~II lines, respectively. Additionally,
$\Teff$, log~g, and [Fe/H] were checked using a grid of
evolutionary tracks from Yi et al. (2001). The stars sit on
the evolutionary tracks corresponding to their mass, age, and metallicity.

\subsection{Influence of inelastic collisions with hydrogen atoms on the non-LTE oxygen abundance depending on atmospheric parameters} 
\label{ostars}

The non-LTE and LTE oxygen abundances for
the sample stars were derived in our earlier study (Zhao et al., 2016).
Here, we redetermined the non-LTE oxygen abundance from the IR 7771-5 \AA\ triplet lines using the
quantum-mechanical data for inelastic collisions with
hydrogen atoms. The mean difference between the
non-LTE abundances from the triplet lines obtained
in this and the previous paper is shown
for each of 46 stars in Fig.~2 as a function of metallicity, effective temperature, and surface gravity.

Just as for the Sun, the application of accurate
data for the sample stars leads to a lower non-LTE
oxygen abundance. The difference in non-LTE abundance ranges from 0.02 to 0.15~dex, depending on the
atmospheric parameters. The difference between the
non-LTE abundances derived with accurate and approximate data diminishes with decreasing metallicity, because, in metal-poor stars, the O~I lines are weak
and the non-LTE corrections are small in absolute value.

\begin{figure}
	\resizebox{160mm}{!}{{\includegraphics{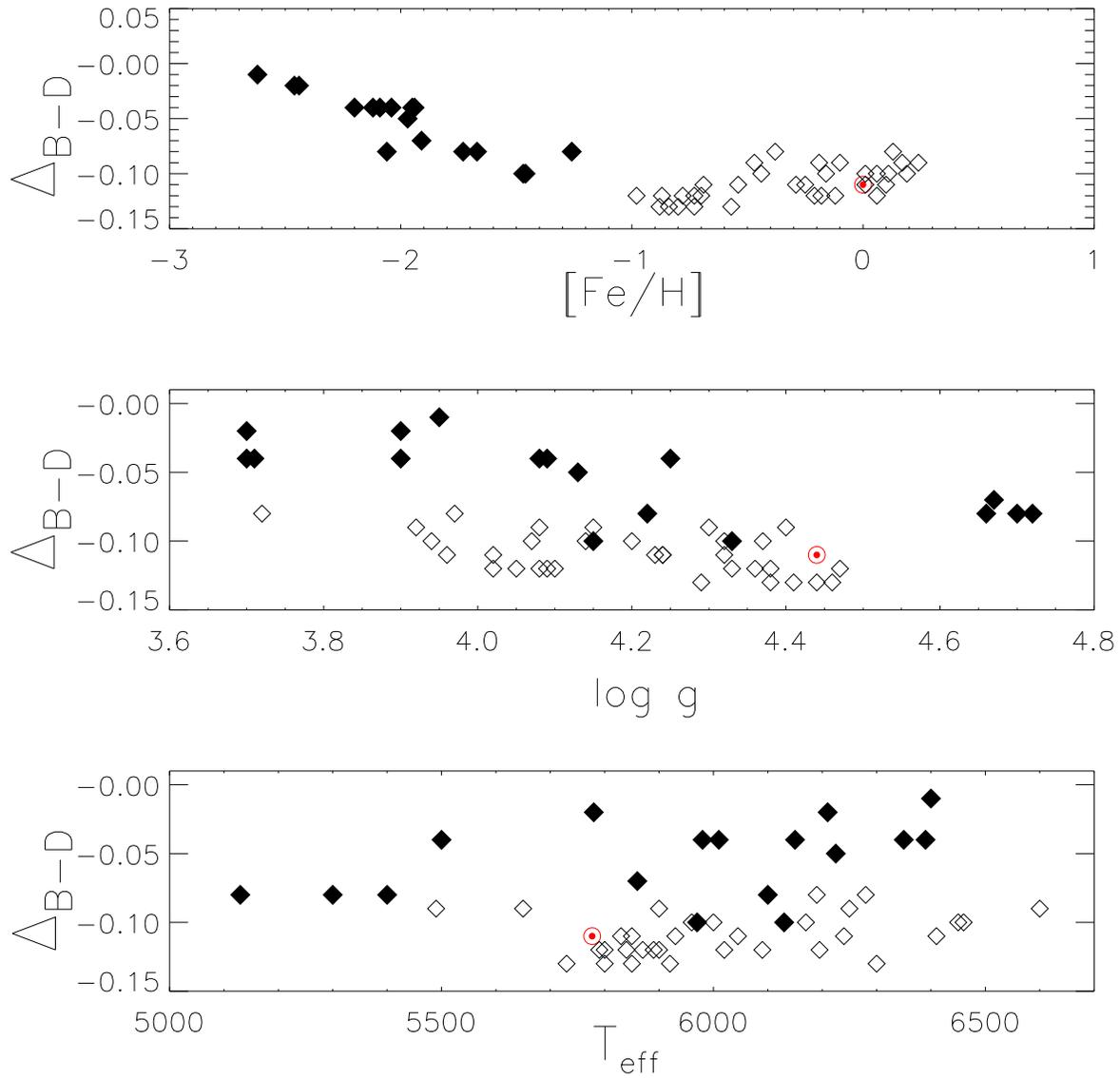}}}
	\caption{
		The mean difference between the non-LTE abundances from the triplet lines derived with accurate
		and approximate data for collisions with hydrogen atoms ($\Delta_{\rm B-D}$) is shown for 46 sample stars and the Sun as a function of
		atmospheric parameters. The filled and open symbols indicate the stars with [Fe/H] < --1 and [Fe/H] > --1, respectively.
		} 
	\label{delta}
\end{figure}

\subsection{The Galactic [O/Fe] trend} 
\label{ofe}

Figure~3 shows the derived [O/Fe] ratio together with those from our previous paper (Zhao et al. 2016). In the metallicity range
--1 < [Fe/H] < 0.2, the [O/Fe] ratio for each star changed within 0.03~dex compared to
the previous results. For these stars, the shift in [O/Fe] is minor, since the change in the  oxygen non-LTE abundance  is approximately the same as that for the Sun. 
For stars with [Fe/H] < --1, the [O/Fe] ratio turns out to be higher than the previous value, and [O/Fe] increasing trend with decreasing [Fe/H] became prominent. This happens, because, for stars with [Fe/H] < --1, the change in the  oxygen non-LTE abundance is smaller in absolute value than those for the Sun and it decreases with decreasing [Fe/H].

Among the modern Galactic chemical evolution models, a model of Kobayashi et al. (2011) fits our previous data for [O/Fe] best of all.
This model predicts a slow decrease
in [O/Fe] from 0.7 to 0.6 as [Fe/H] increases from --3
to --1 and then a sharp drop in [O/Fe] down to 0.1 at [Fe/H] =
--0.17 (the values were recalculated by taking into
account the difference between our previous solar oxygen and
iron abundances and those adopted by Kobayashi et al. (2011)). 
The revision of the oxygen abundance
led to an increase in [O/Fe] with decreasing metallicity at [Fe/H] < --1 instead of a plateau at [O/Fe] = 0.6, as derived in our previous study. 
The rising trend in [O/Fe] with decreasing [Fe/H] < --1 can be qualitatively described by the model of Kobayashi et al. (2011), where rapidly rotating metal-poor massive stars were taken into account. 
Including rapidly rotating massive stars improves significantly the description of the observed
behavior of [C/Fe] and [N/Fe] at [Fe/H] < --1.5, but
does not affect the behavior of heavier elements.
The model  predicts oxygen abundance of $\eps$ = 8.93 at [Fe/H] = 0 and fit the determinations of Anders and Grevesse (1989) for the Sun.
In absolute scale, this model  fits well the data from Zhao et al. 2016 at [Fe/H] < --1, while the predicted  [O/Fe] is larger than the observed at --1 < [Fe/H] < 0.2. The current revision in solar and stellar oxygen abundance leads to a discrepancy between model predictions and observations throughout all metallicity regimes.

\begin{figure}[htbp]
	\resizebox{140mm}{!}{{\includegraphics{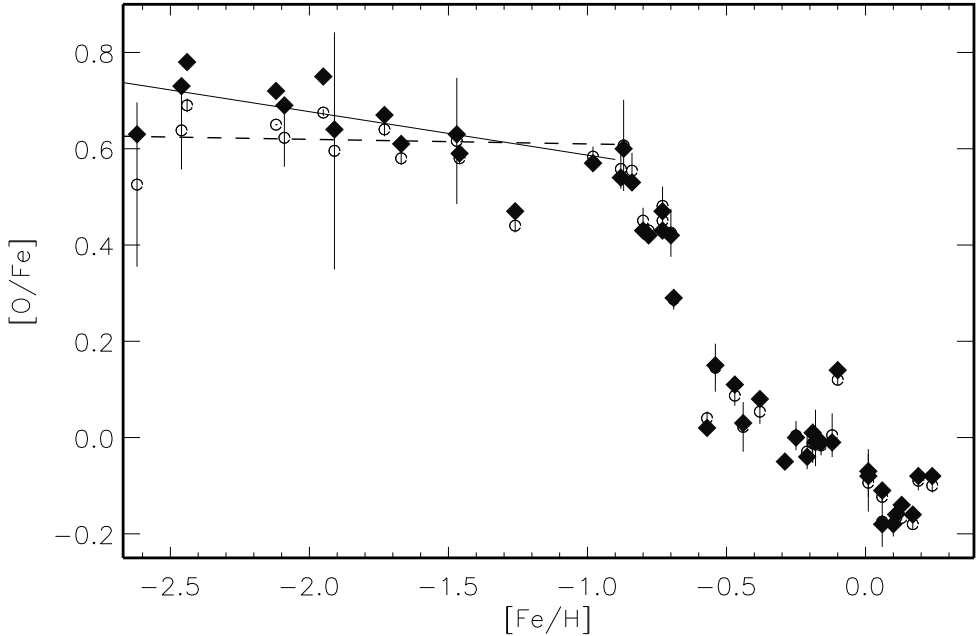}}}
	\caption{
The [O/Fe] ratio derived in non-LTE with the accurate (diamonds) and approximate (circles) rates of inelastic  collisions
with hydrogen atoms is shown for 46 sample stars as a function of [Fe/H]. The straight lines [O/Fe] = 0.60--0.01[Fe/H]
(dashed) and [O/Fe] = 0.50--0.09[Fe/H] (solid) indicate the linear interpolation result in the range --2.6 < [Fe/H] < --1.3.
We excluded the star HD~103095 with [Fe/H] = --1.26 when calculating the slopes of the straight lines, because it belongs to
the type of low-$\alpha$ stars.		
		 } 
	\label{ofe_figure}
\end{figure}

\section{Conclusions} 
\label{conclusions}

We performed non-LTE calculations for O~I using the data from Barklem (2018) for collisions with
hydrogen atoms. 
The non-LTE abundances from the O~I lines were determined for the Sun and 46 FG stars in a wide metallicity range, $-2.6 <$ [Fe/H] $< 0.2$. 

The application of accurate data leads to a strengthening of the departures from LTE and a decrease
in the abundance from the IR lines. The largest changes
in the non-LTE abundance we found for the IR
7771-5 \AA\ triplet lines. For the Sun, the
non-LTE abundance from these lines decreases by
0.11 dex compared to what is obtained with approximate data for collisions with hydrogen atoms. Our
new calculations do not affect the non-LTE abundance from weak lines in the visible range (6158, 6300 \AA).

For the Sun, we obtained the mean non-LTE
oxygen abundance $\eps$(O) = 8.69 $\pm$ 0.08, which is
lower than that inferred in non-LTE with approximate
data for collisions with hydrogen atoms by 0.07 dex.
This value is even farther from what is required to
reconcile the theoretical and observed density and
sound speed profiles.

For 46 stars of the sample, the changes in the
non-LTE abundance vary from 0.02 to 0.13 dex, depending on the atmospheric parameters. The change
in the solar oxygen abundance and the oxygen abundance for individual stars led to a change in the
behavior of the Galactic [O/Fe] trend with [Fe/H]
in the range --2.6 < [Fe/H] < --1: the increase in
[O/Fe] with decreasing [Fe/H] became more noticeable. For stars with --1 < [Fe/H] < 0.2, the changes
in the non-LTE abundance for the sample stars are
close to what was obtained for the Sun. Therefore,
the [O/Fe] ration barely changed.
The refined [O/Fe] trend  can be used for testing of
Galactic chemical evolution models.

{\bf Acknowledgments:}
This work was supported by the Russian Science
Foundation (grant no. 17-13-01144). We are grateful
to K. Fuhrmann for the spectra and O. Kochukhov for
the binmag code. We made use the VALD and MARCS databases.
This study is based on observations made with ESO Telescopes at the La Silla Paranal Observatory
and at the Canada-France-Hawaii Telescope (CFHT), which is operated
by the National Research Council (NRC) of Canada, the Institut National des Science de l'Univers of the Centre National de
la Recherche Scientifique (CNRS) of France, and the University
of Hawaii. The operations at the CFHT are conducted with care
and respect from the summit of Maunakea, which is a significant
cultural and historic site.
This work has made use of data from the European Space Agency (ESA) mission
{\it Gaia} 
(https://www.cosmos.esa.int/gaia),
 processed by the {\it Gaia}
Data Processing and Analysis Consortium (DPAC,
https://www.cosmos.esa.int/web/gaia/dpac/consortium). Funding for the DPAC
has been provided by national institutions, in particular the institutions
participating in the {\it Gaia} Multilateral Agreement.


%

\end{document}